\documentstyle[epsf,dcolumn]{mn}
\begin{document}
% to prevent inclusion of figures, uncomment the next line
%\epsfdrafttrue
\newcolumntype{d}[1]{D{.}{.}{#1}}
\newcommand{\twiddles}{\sim}
\newcommand{\etal}{{et al}\/.}
\newcommand{\uv}{{\it uv}}
\def\Ssin#1C#2 {#1C\,#2}
\def\Ss#1{\Ssin#1 }
\def\afterpage#1{}
\def\dgr{^\circ}
\title[FR\,II radio galaxies with $z < 0.3$ -- II]{FR\,II radio galaxies
with $\bmath{z < 0.3}$ -- II. Beaming and unification}
\author[M.J.~Hardcastle \etal]{M.J.~Hardcastle$^{1,2}$\thanks{E-mail {\it M.Hardcastle}@{\it bristol.ac.uk}},
P.~Alexander$^2$, G.G.~Pooley$^2$ and J.M.~Riley$^2$\\
$^1$ Department of Physics, University of Bristol,
Royal Fort, Tyndall Avenue, Bristol BS8 1TL\\
$^2$ Mullard
Radio Astronomy Observatory, Cavendish Laboratory, Madingley Road,
Cambridge, CB3 0HE\\}
%\maketitle
% now do it all again
{\par
 \begingroup
\twocolumn[
\vspace*{17pt}\raggedright \sloppy\huge\bf {FR\,II radio galaxies
with $\bmath{z < 0.3}$ -- II. Beaming and unification}\vskip 23pt
\LARGE
M.J.~Hardcastle$^{1,2}$\footnotemark, P.~Alexander$^{2}$,
G.G.~Pooley$^2$ and J.M.~Riley$^2$\par
\small{\it $^1$ Department of Physics, University of Bristol, Royal Fort, Tyndall
Avenue, Bristol BS8 1TL\\
$^2$ Mullard Radio Astronomy Observatory, Cavendish
Laboratory, Madingley Road, Cambridge, CB3 0HE}\par
\vskip 22pt
\today\par\vskip 22pt
]
 \thispagestyle{titlepage}
 \endgroup
}

\begin{abstract}
In a previous paper we presented measurements of the properties of
jets and cores in a large sample of FRII radio galaxies with
$z<0.3$. Here we test, by means of Monte Carlo simulations, the
consistency of those data with models in which the prominences of
cores and jets are determined by relativistic beaming. We conclude
that relativistic beaming is needed to explain the relationships
between core and jet prominences, and that speeds between 0.5 and
0.7$c$ on kpc scales provide the best fits to the data.
\end{abstract}

\footnotetext{E-mail {\it M.Hardcastle}@{\it bristol.ac.uk}}

\section{Introduction}
\label{intro}

There is strong and widely accepted evidence for relativistic bulk
speeds on parsec scales in radio galaxies and quasars. In the most
extreme objects the observed superluminal motions, the rapid
variability and consequent high brightness temperatures, and the
absence of very strong inverse-Compton emission in the X-ray seem to
require relativistic bulk speeds with $\gamma \ga 2$; it seems
reasonable to assume that such high bulk Lorentz factors are present
close to the active nucleus in all extragalactic radio
sources. Evidence that relativistic speeds persist on kiloparsec
scales is less conclusive, but in FRII objects a one-sided jet on
parsec scales usually connects to a one-sided jet on kiloparsec scales,
and the result of Laing (1988) and Garrington \etal\ (1988) that the
brighter jet in an FRII radio source tends to lie in the less
depolarized lobe is very hard to explain unless relativistic beaming
is still important on these scales.

Relativistic beaming is an essential ingredient in unified models for
powerful radio galaxies and quasars (e.g.\ Barthel 1989); on the
assumption that quasars are radio galaxies oriented at less than some
critical angle to the line of sight, relativistic bulk speeds in the
core and jet imply that the cores and jets of quasars should be more
prominent relative to the extended emission and that their jets should
be more one-sided. Until recently, the lack of detected jets in FRII
radio galaxies, while giving qualitative support to this model, meant
that it could not be tested quantitatively.

In a previous paper (Hardcastle \etal\ 1998; paper I) we discussed the
properties of radio jets and cores in a large sample of 3CR FRII radio
galaxies with $z<0.3$ which had mostly been imaged with the NRAO VLA
at high resolution and sensitivity. Since there are no quasars in the
parent samples at this low redshift, there is necessarily a population
of galaxies in our sample aligned at small angles to the line of
sight. We suggested that our results were consistent with a modified
version of unified models in which broad-line radio galaxies (BLRG)
were the aligned objects and the low-power counterparts of quasars;
the jets and cores of the BLRG were systematically more prominent than
those of narrow-line radio galaxies (NLRG), and the relative numbers
of NLRG and BLRG were consistent with their being divided by a
critical angle to the line of sight of $\sim 45\dgr$, similar to that
found by Barthel (1989) from the relative numbers of higher-redshift
3CR radio galaxies and quasars. The radio properties of the BLRG in
our sample were similar to those of the FRII quasars studied by Bridle
\etal\ (1994). Assuming that relativistic beaming is the dominant
factor determining the prominences of the cores, the relationships
between the properties of cores and jets implied that the jets were
beamed, with characteristic speeds in the jets lower than those in the
cores, consistent with the results of Bridle \etal\ (1994). In this
paper we attempt to make this conclusion quantitative by fitting
empirical models of relativistic beaming effects to the data derived
from the sample.

The observed flux density $S_{\rm obs}$ of a jet with speed $v = \beta
c$ and Lorentz factor $\gamma = (1-\beta^2)^{-{1\over 2}}$, emitting
isotropically in its rest frame, is related to the flux density that
would be observed if it were at rest ($S_{\rm rest}$) by the relation
\begin{equation}
S_{\rm obs} = S_{\rm rest}[\gamma(1 - \beta \cos
\theta)]^{-(m+\alpha)}\label{db}
\end{equation}
(Ryle and Longair 1967) where $\theta$ is the angle made by the
velocity vector with the line of sight, $\alpha$ is the spectral index
(defined throughout in the sense $S \propto \nu^{-\alpha}$), and $m$
is a constant reflecting the geometry of the beamed component
(e.g. Scheuer \& Readhead 1979); we use $m=2$, the value appropriate
for a continuous jet. Because in practice the material in beams is
expected to have a range of speeds, care is necessary in interpreting
values of $\gamma$ and $\beta$ deduced from this expression as
straightforward physical quantities; they nevertheless provide a means
of characterising the physical parameters in the beam\footnote{As in
paper I, we try to follow Hughes \& Miller (1991) in distinguishing
between the `beam', the stream of particles supplying the hot spots
and lobes with energy, and the `jet', its observable manifestation.}.

The approach to modelling relativistic beaming that we adopt has been
widely used in testing unified models of different classes of radio
sources (e.g.\ Scheuer \& Readhead 1979; Urry \& Schafer 1984; Urry,
Padovani \& Stickel 1991; Morganti \etal\ 1995). The prominence of a
radio component is defined as the ratio of the flux density of the
component to the flux density of the {\it extended} emission (which we
define as the total flux density of the source minus any detected
emission from the core and jets). It is assumed that the extended
emission is unaffected by beaming; the prominence $p'$ of a beamed
component is then taken to be given by the product of some intrinsic
prominence $p$ and the Doppler beaming factor given by equation
\ref{db};
\begin{equation}
p' = p [\gamma(1 - \beta \cos
\theta)]^{-(m+\alpha)}
\label{promdb}
\end{equation}
To see whether observed data are consistent with a particular set of
assumptions about the parameters of such a model, we generate large
numbers of simulated sources by Monte Carlo methods (assuming an isotropic
distribution of angles to the line of sight) and compare the resulting
distributions with the observations. By varying the unknown parameters
($p$ and $\beta$ in this simple model) and examining the
goodness of the fit, it is possible to find the values of the
parameters that produce simulated datasets that, on average, best
reproduce the data (i.e., have the highest probability, on some
statistical test, of being drawn from the same parent
population). Depending on the fitting statistic used and the space
over which it is maximised, this approach may or may not be formally
identical to a maximum-likelihood analysis; but in any event we expect
it to provide insights into the regions of parameter space suggested
by the data.

In the rest of the paper we apply this simple technique to the data of
paper I.

\section{The data}
\label{data}

The sample of paper I consisted of 50 radio galaxies from the samples
of Laing, Riley \& Longair (1983) and the 3CR catalogue (Spinrad
\etal\ 1991). All had redshifts less than $0.3$ and radio power at 178
MHz greater than $1.5 \times 10^{25}$ W Hz$^{-1}$ sr$^{-1}$, above the
canonical FRI-FRII break (Fanaroff \& Riley 1974) although a few had
structures which resembled those found in FRIs (e.g. apparently
dissipative jets). The reader is referred to paper I for details of
the selection criteria. In what follows we assume, as argued in that
paper, that the selection was not strongly biased in ways that might
affect unified models; in particular, that objects at all angles to
the line of sight are represented in the sample.

High-resolution radio maps were available to us for 44 of the 50
objects, and from them we were able to measure the flux densities of
cores and jets and counterjets, or to set upper limits on these
quantities where the components were not detected. These quantities,
are reproduced from paper I in Table \ref{numbers}.  As argued by
Bridle \etal\ (1994), the {\it straight} component of the jet is the
only part in which relativistic beaming effects can usefully be
studied, and throughout this paper `jet flux density' and `jet prominence'
will be used to refer to the flux density and prominence of the straight
component of the (brighter) jet only, as defined in paper I. We do not
expect to introduce any bias into the data by considering the straight
jet only. The prominences of the core and straight jet are also
tabulated. If more than one jet is detected in a given source, the
prominence of the brighter jet is used; if only one jet is detected,
we tabulate its prominence; if no jets are detected, the larger of the
two upper limits on jet flux density is used to derive an upper limit
on jet prominence. The errors quoted for the fluxes of detected jets
are intended to reflect the uncertainty in the separation of the jet
from its background, as discussed in paper I, and are not formal
errors. Jets or possible jets were detected in 33 of these 44 objects;
radio cores were detected in all but 2.

\begin{figure*}
\begin{center}
\leavevmode
\epsfysize=15cm\epsfbox{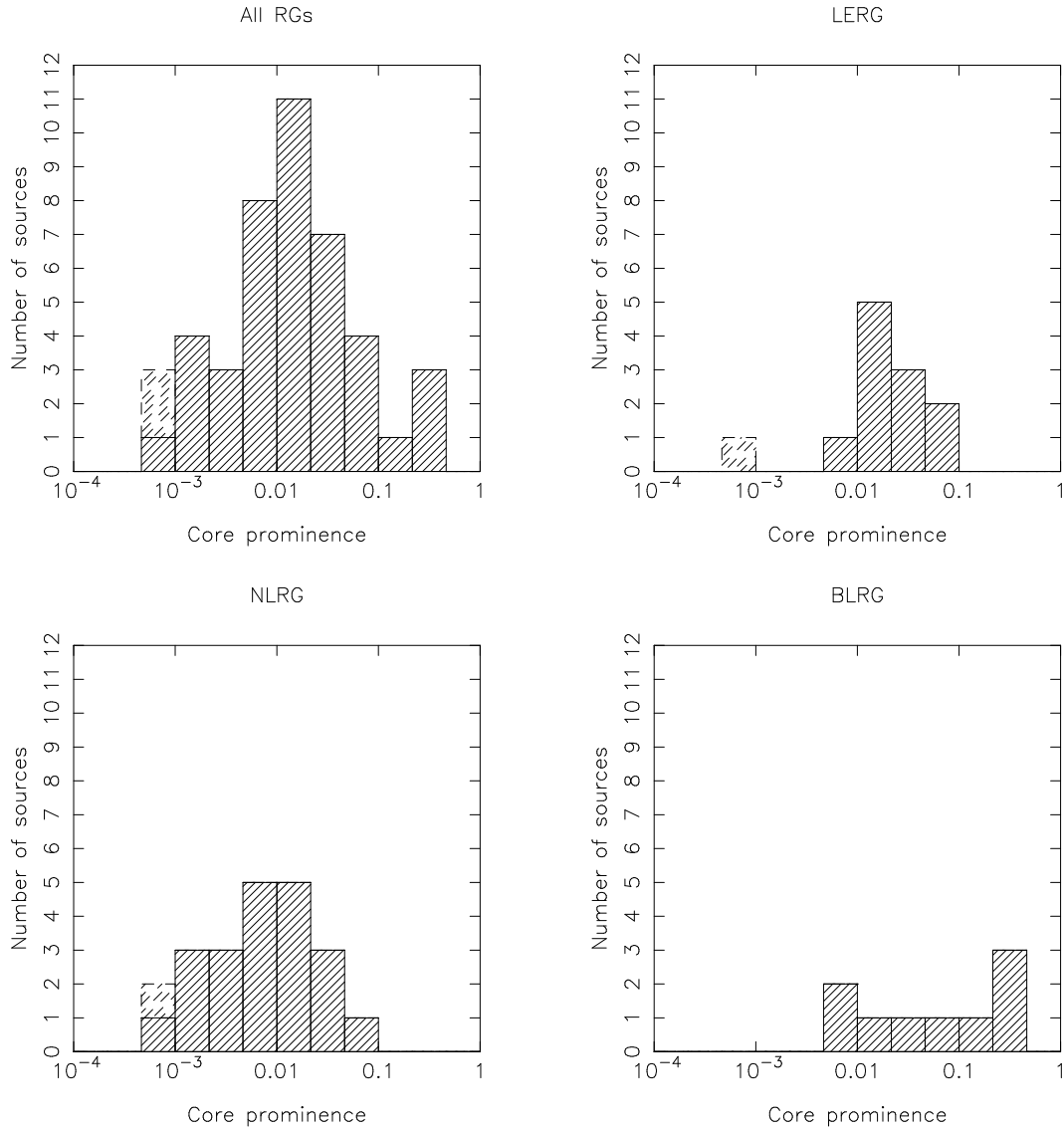}
\caption{Core prominences of the objects in the sample classed by
emission-line type. Dashed shading indicates an upper limit.}
\label{coreprom}
\end{center}
\end{figure*}

\begin{figure*}
\begin{center}
\leavevmode
\epsfysize=15cm\epsfbox{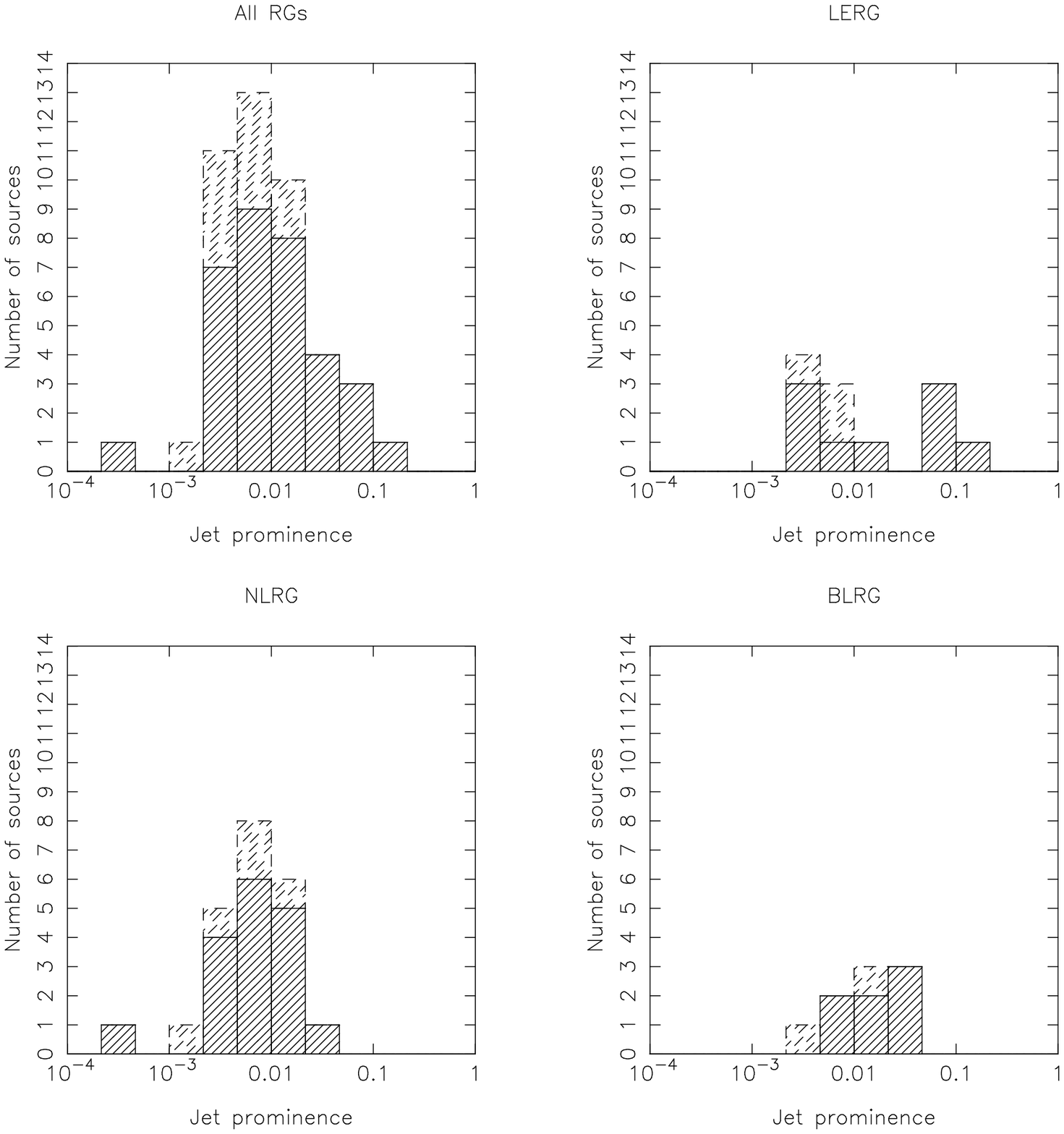}
\caption{Jet prominences of the objects in the sample classed by
emission-line type. Dashed shading indicates an upper limit.}
\label{jetprom}
\end{center}
\end{figure*}

We also tabulate the optical emission-line classes of the sources. We
argued in paper I that there is a real difference, in this sample,
between the properties of those sources with low-excitation emission
lines (low-excitation radio galaxies, LERG) and those with stronger
emission lines (the narrow-line radio galaxies, NLRG, and broad-line
radio galaxies, BLRG). The prominences of cores and jets for the
different classes of object are plotted in Figures \ref{coreprom} and
\ref{jetprom}. Fig.\ \ref{jetprom} shows that the most prominent jets
are actually those of a subset of the LERG; in paper I we suggested
that these were sources with particularly dissipative beams. For this
reason we restrict ourselves in what follows to the sub-sample of 31
high-excitation objects.

\section{Model fits}
\label{fits}

Both the distributions of core and jet prominence (plotted in Figs
\ref{coreprom} and \ref{jetprom}) and the {\it} relationship between
these quantities (Fig.\ \ref{corejet}) contain information about the
best-fit beaming parameters. It is clear that the distributions could
individually be consistent with a relativistic beaming model while
exhibiting some relationship (for example, a core-jet prominence
anticorrelation) that effectively ruled such a model out. In paper I
we showed that both the distributions of these quantities and the
positive correlation between them were {\it qualitatively} consistent
with a model in which both cores and jets are relativistically beamed,
and with unified models in which the BLRG were the objects more
closely aligned with the line of sight. Traditionally the procedure
outlined in Section \ref{intro} has been applied to a single
distribution, e.g. to core prominences (Morganti \etal\ 1995) or jet
sidednesses (Wardle \& Aaron 1997). We begin our quantitative approach
to the problem in that way; later we discuss the implications of the
core-jet relationship.

A serious problem in the data analysis arises because of the number of
upper limits in the distributions, particularly in the jet
prominences. Whether a jet is apparent in a particular map or not
depends not just on the sensitivity of the observations, but also on
the presence or absence of confusing features in the lobe; it is
therefore difficult to reproduce the non-detections in the data in
numerical simulation. We believe that the upper limits we determine
are probably not far above the true values, but in what follows we
shall try to assess the effects of the upper limits on the results of
the fits.

In order to apply the model discussed in section \ref{intro} to a
large number of sources, it is necessary to make some assumptions
about the intrinsic prominence $p$ of the component of interest. The
simplest assumption, adopted by Urry \& Schafer (1984) and
subsequently followed by other authors, is that the intrinsic
prominence of jets or cores is fixed (equivalently, that the intrinsic
luminosity of jets or cores is a fixed fraction of the total
luminosity of the radio source) so that all sources have the same
intrinsic prominence for a given component. This is a reasonable
assumption so long as relativistic beaming is expected to be the
dominant effect in determining the prominence of a component. In the
case of radio galaxies, it seems likely that the environment of, and
`weather' in, a particular radio source can strongly affect the
efficiency of jets and cores, and it is certainly true that the
environment will affect the {\it extended} luminosity of a source. The
normalisation by extended flux density removes much of the dependence
of the prominences of jet and core on the underlying power of the
radio source (`beam luminosity') but cannot eliminate it entirely. All
of these factors affect component prominences, in general by
introducing additional scatter into the distribution of observed
prominences. In our analysis we begin by making the simplest
assumptions possible, and go on to discuss the effects of relaxing
them.

The statistical test we use for comparing real and simulated data
throughout the paper is the Kolmogorov-Smirnov (K-S) two-sample test;
the statistic provided by this test can be converted to a probability
that the two samples compared are drawn from the same parent
population [see e.g.\ Press \etal\ (1992) for a discussion of the
properties and implementation of this test]. We prefer this test to
the $\chi^2$ test because of the loss of power involved in binning our
small sample; however, wherever we compare two one-dimensional
distributions in the following sections we find that the results when
a $\chi^2$ test is used are consistent with those of the K-S test.

As discussed in section \ref{data}, we believe that the data are not
biased with respect to their orientation to the line of sight; this
allows us to neglect the details of unified models and to treat the 31
NLRG and BLRG as a homogeneous sample. The clear differences,
discussed in paper I, between the core and jet prominences of the BLRG
and NLRG motivate the belief that relativistic velocities are involved
on both scales; but it would also in principle be possible to use the
presence or absence of broad lines as an additional source of
information on the inclination of sources in the sample when carrying
out model fits. We have chosen not to do this for two reasons: firstly
because the presence or absence of broad lines in these objects is
strongly dependent on the quality of available optical observations
(see e.g. Laing \etal\ 1994) and because this raises the question of
the status of objects in the sample, like 3C\,234, with polarized
broad lines; secondly because to generate comparison samples requires
a knowledge of the critical angle separating the two populations,
which, even if such a simple unified model adequately represents the
true situation, is not well constrained, given the small sample size,
by the existing data on numbers of sources, supposing the numbers
themselves to be accurate. We therefore restrict ourselves, in this
section, to noting the degree of agreement between the data and the
predictions of a simple unified model (with critical angle $45\dgr$)
based on our best-fit models.

Throughout the rest of the paper a subscript c denotes a quantity related to
the core of a source and a subscript j denotes one related to the jet.

\subsection{Cores}
\label{cores}

Only one of the core prominences for the high-excitation objects is an
upper limit (that for 3C\,153) and therefore we ignore the problem of
upper limits in this part of the analysis, treating the upper limit on
3C\,153's core prominence as a measurement. We assume that the
spectral index of cores is zero. Because cores are the unresolved
bases of jets, which we presume to be intrinsically two-sided, the
correct model to use for our sample, where there are angles to the
line of sight $\theta$ close to $90\dgr$, is

\begin{equation}
p'_c = p_c \left\{ [\gamma(1 - \beta_c \cos
\theta)]^{-(m+\alpha)} + [\gamma(1 + \beta_c \cos
\theta)]^{-(m+\alpha)} \right \}
\label{corepromdb}
\end{equation}

--- this is identical to equation \ref{promdb} at small $\theta$ and
large $\beta_c$, where the second term is negligible in comparison to
the first.

\begin{figure*}
\begin{center}
\leavevmode
\vbox{\epsfxsize=13cm
\epsfbox{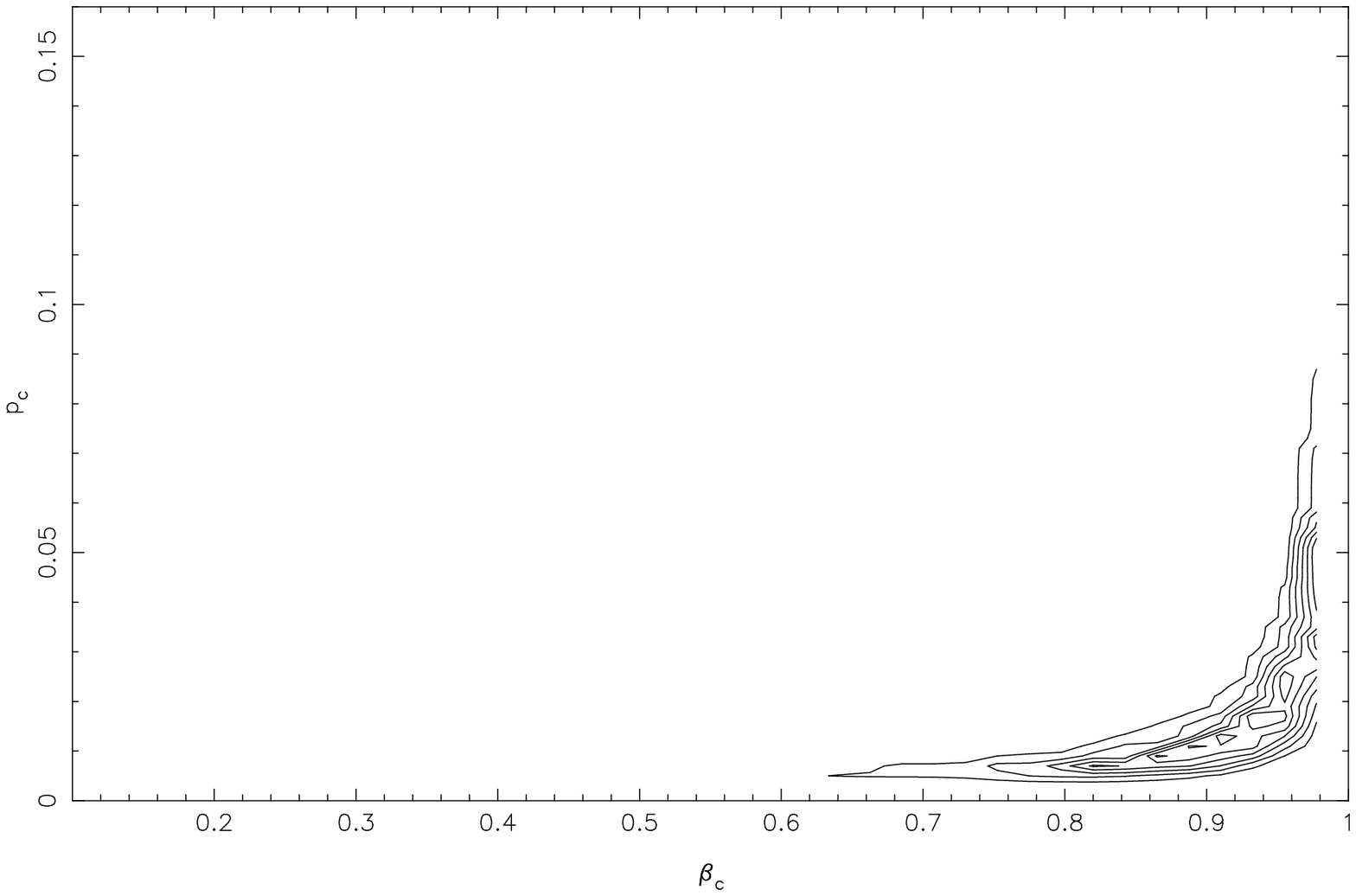}
}
\vbox{\vskip 10pt\epsfxsize=13cm
\epsfbox{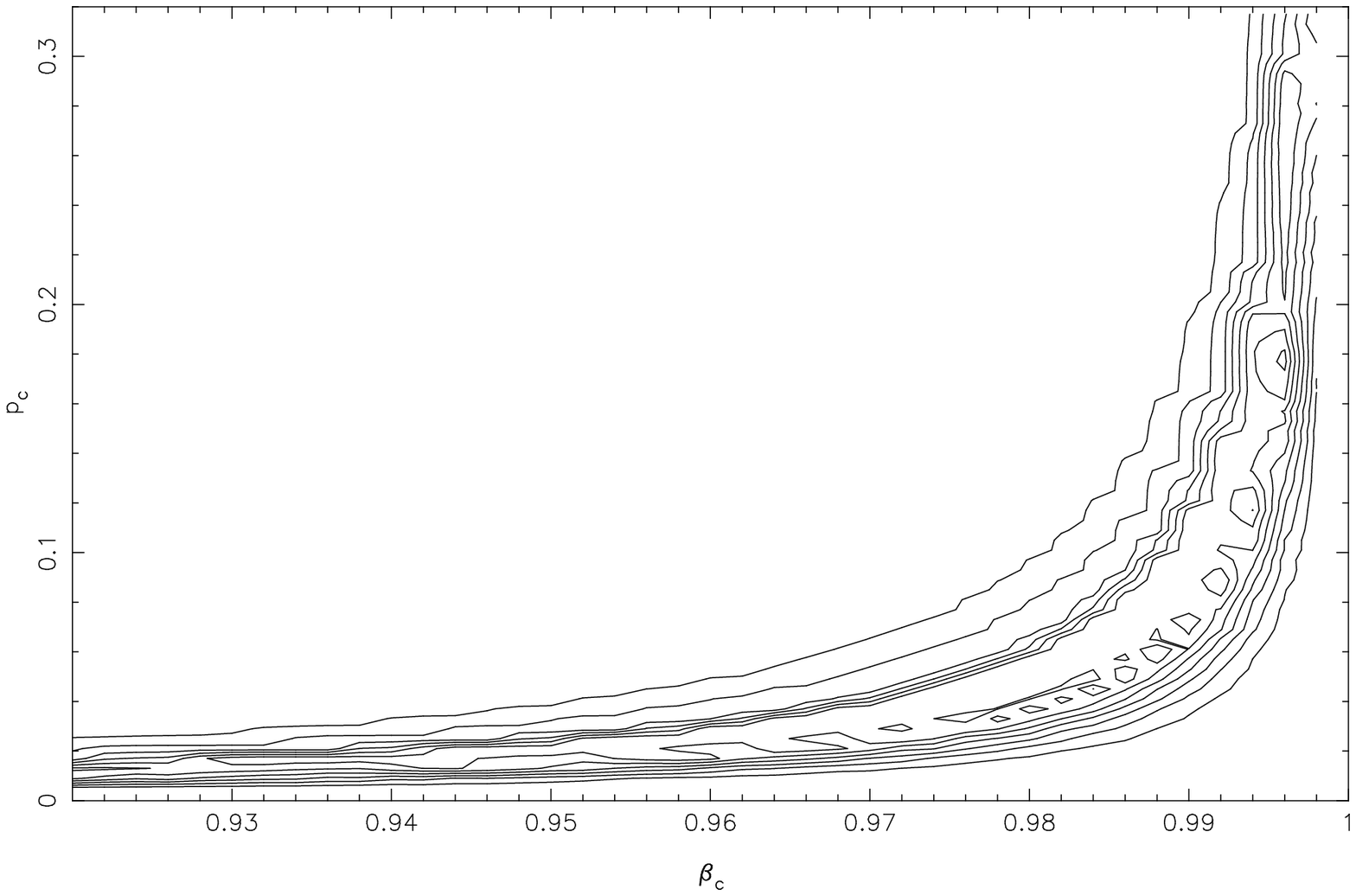}}
\caption{Kolmogorov-Smirnov probability contours for model core
prominence data fitted to the prominences of BLRG and NLRG. Beaming
speed $\beta_c$ is plotted on the $x$-axis. Intrinsic prominence $p_c$
is plotted on the $y$-axis. Contours at K-S probabilities of $0.05,
0.1, 0.15\dots, 0.40$. Above, $\beta_c$ ranges from $0.1$
to $0.98$; below, as above but showing the region with $0.92 <
\beta_c < 0.998$.}
\label{core-large}
\end{center}
\end{figure*}

Fig.\ \ref{coreprom} shows the distribution of core prominences in the
NLRG and BLRG to be very broad. It is therefore not surprising that
when models of the form described in section \ref{intro}, using the
prominence described by equation \ref{corepromdb}, are fitted to the
data only those with large values of $\beta_c$ can reproduce the data
well (Fig.\ \ref{core-large}). The data do not put an upper constraint
on $\beta_c$, but they are best reproduced with $\beta_c > 0.95$
(corresponding to $\gamma_c \ga 3$). The lack of an upper constraint
is due to the relatively few highly beamed objects in a sample without
orientation bias. VLBI information constraining the speeds in the
cores is only available for two of our objects: superluminal motion in
the core of 3C\,111 constrains $\gamma \ga 6.9$ (Preuss \etal\ 1990)
and in 3C\,390.3 $\gamma \ga 3.6$ (Alef \etal\ 1996). Our best-fit
Lorentz factors are consistent with these observations. For $\beta_c =
0.98$, under our assumed unified model with a critical angle of
$45\dgr$, we would expect the median BLRG core prominence in a sample
of 31 objects to be about 18 times the median NLRG core prominence,
which is larger than the observed ratio of about 8; however, the ratio
observed in our data is not inconsistent with this value of the
critical angle at the 95 per cent confidence level for any value of
$\beta_c$ allowed by the distribution of core prominences, though it
is definitely inconsistent with a critical angle $\la 40\dgr$ unless
$\beta_c <0.98$.

In paper I we discussed a possible positive correlation, only apparent
in the narrow-line objects, between core prominence and apparent
source linear size. Regression [performed using the Buckley-James
method, implemented in the survival analysis software package {\sc
asurv} Rev.~1.1; LaValley, Isobe \& Feigelson (1992)] suggests a
non-linear relationship between the two quantities, $p_c' \propto (L
\sin \theta)^{1.6 \pm 0.4}$. This might be attributed to a
proportionality between unprojected source length $L$ and {\it intrinsic}
core prominence $p_c$; such an effect would tend to be seen
preferentially in the narrow-line objects, since it would be diluted
by projection and beaming effects in objects at small angles to the
line of sight. We may consider two hypotheses:
\begin{enumerate}
\item Intrinsic core prominence is independent of source length;
prominences only depend on $\beta_c$ and $\theta$ as in equation
\ref{promdb}, and the observed correlation is coincidental.
\item Intrinsic core prominence is dependent on source
length; prominences depend on $\beta_c$, $\theta$ and $L$, the
unprojected source length.
\end{enumerate}
In the latter case the core prominence of a source is given by
\begin{equation}
p'_{\rm c} = k_{\rm c,l} L^a \left\{ [\gamma(1 - \beta_c \cos
\theta)]^{-(m+\alpha)} + [\gamma(1 + \beta_c \cos
\theta)]^{-(m+\alpha)} \right\}
\label{promdb2}
\end{equation}
where $a$ governs the relationship between $L$ and $p'_c$. The effects of
projection and beaming will tend to flatten the power-law slope of the
observed relationship compared to the true one. The magnitude of this
effect, observed in simulated datasets, depends on $\beta_c$ and the
unification angle, but for plausible values of both we expect the
observed slope to be flatter than the true slope by about 0.2, so that
the true slope $a$ is likely to be about $1.8 \pm 0.4$. For
simplicity, therefore, we choose $a=2$, a value consistent with the
data. $k_{\rm c,l}$ is the constant of proportionality in this
relationship and clearly (for this choice of $a$) has units of
kpc$^{-2}$.

To use Monte Carlo methods to test these models against the data, we
first need to model the distribution of unprojected source lengths in
the parent population. Little is known about this distribution. The
simple empirical form suggested by Kapahi (1976) is not a particularly
good fit to our data, as it overpredicts the number of small
sources. Simulation shows that the data are moderately well described
by a simple normal distribution of unprojected length (mean 560 kpc,
standard deviation 230 kpc) with negative lengths being discarded. We
use simulated unprojected lengths drawn from this distribution to test
the hypotheses.

\begin{figure*}
\begin{center}
\leavevmode
\epsfxsize=13cm
\epsfbox{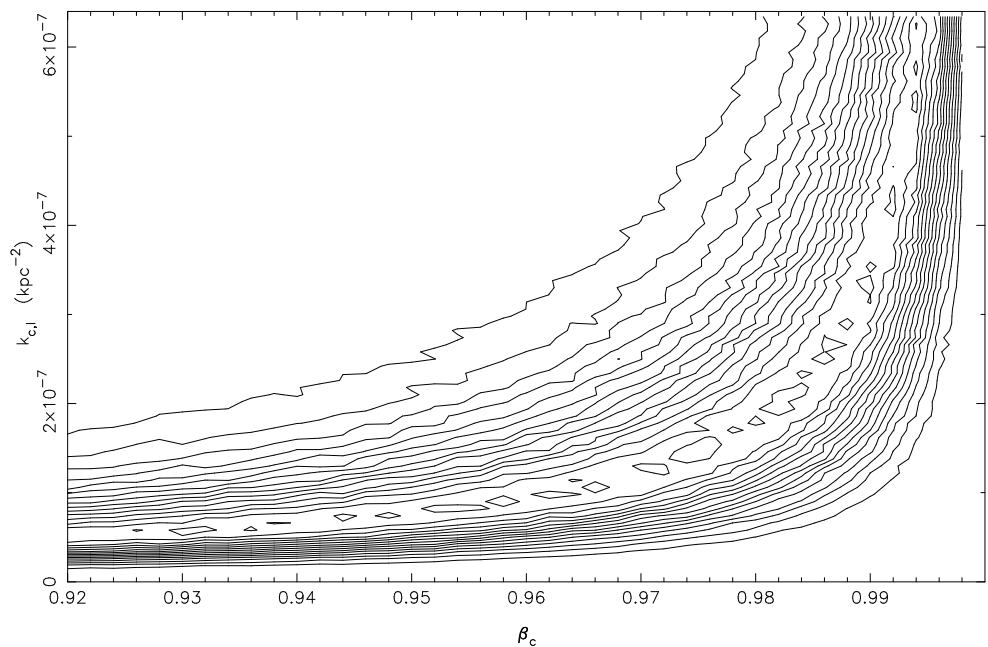}
\caption{Kolmogorov-Smirnov probability contours for model core
prominence data fitted to the prominences of BLRG and NLRG, including a
dependence of intrinsic prominence on source length as discussed in
the text. Beaming speed $\beta_c$ is plotted on the $x$-axis, and
varies between $0.92$ and $0.998$; the constant relating length to
intrinsic prominence, $k_{\rm c,l}$ (kpc $^{-2}$), is plotted on the
$y$-axis. Contours at K-S probabilities of $0.05, 0.1, 0.15\dots, 0.75$.}
\label{newcore}
\end{center}
\end{figure*}

We first repeated the core prominence simulations to find the best-fit
values of $k_{\rm c,l}$ and $\beta_c$ for model (ii).  The results of
the simulations for $\beta_c$ are not strongly affected by the
inclusion of a length dependence, presumably because the distribution
of core prominences in the data is much broader than that of
lengths. Lower values of $\beta_c$ become slightly more probable, as
we would expect (Fig. \ref{newcore}). There is again no constraint on
the maximum value of the beaming speed $\beta_c$, so we choose a
representative value of $\beta_c=0.98$, corresponding to $k_{\rm c,l}
\sim 2.5 \times 10^{-7}$ kpc$^{-2}$. Using these values, we were able
to generate pairs of (projected length, apparent core prominence)
values to match those in the real dataset. We can then answer the
question: how likely is it under the two hypotheses that 22 NLRG cores
would produce the significant positive correlation seen on a Spearman
Rank test in the real data? Assuming that a critical angle of $45\dgr$
to the line of sight divides broad- and narrow-line objects, and
setting the significance level for the one-tailed Spearman-Rank test
at 99\%, we find that under model (i) only 0.1\% of simulated samples
could produce such a result, whereas under model (ii) approximately
94\% of samples can do so. Model (i) may be rejected at a high level
of significance, but model (ii) is consistent with the data. The
physical interpretation of this result is unclear.

\subsection{Jets}
\label{jets}

\begin{figure*}
\begin{center}
\leavevmode
\epsfxsize=13cm
\epsfbox{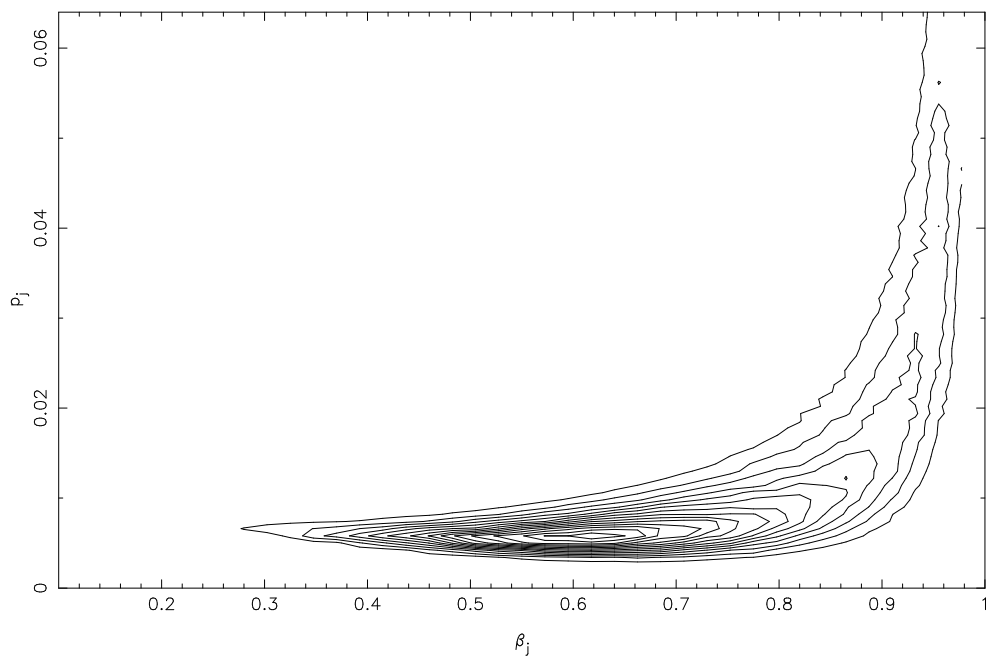}
\caption{Kolmogorov-Smirnov probability contours for model jet
prominence data fitted to the prominences of BLRG and NLRG. Beaming
speed $\beta_j$ is plotted on the $x$-axis, and varies between $0.1$
and $0.98$; intrinsic prominence $p_{\rm j}$ is plotted on the
$y$-axis. Contours at K-S probabilities of $0.05, 0.1, 0.15\dots, 0.70$.}
\label{jetks}
\end{center}
\end{figure*}

We begin by treating the upper limits on jet prominence as
measurements and considering only beaming to be responsible for the
distribution of observed jet prominences. We take the spectral index
of a jet to be $0.8$; the analysis is not significantly affected if
this number is altered. The results of the simulations are plotted in
Fig.\ \ref{jetks}. There is a clear best-fit value, corresponding to
$\beta_j = 0.62$, $p_j=0.005$, although again a wide range of
parameters is formally acceptable. Assigning a confidence range in the
standard way, we find that $\beta_j = 0.62^{+0.04}_{-0.1}$ at the 90
per cent confidence level. For a unified model with critical angle
$45\dgr$, we would expect the median BLRG jet with this value of
$\beta_j$ to be about 4 times more prominent than the median NLRG jet;
this appears consistent with the distribution seen in Fig.\
\ref{jetprom}, though it is somewhat higher than the ratio ($\sim 2$)
of the medians quoted in paper I, where the median jet prominence was
calculated on the assumption that all the upper limits fell below the
median. The data are better fitted by smaller values of $\beta_j$
and/or larger critical angles.

If we were able to replace our upper limits with true measurements, we
would expect the distribution of jet prominences to broaden, so that
the best-fit values of $\beta_j$ would increase. Although we do not know
how far below the upper limits the true measurements lie, we can
assess the effect that they have on the results of the fits by
allowing them to vary while the fits are being performed. We therefore
ran a set of fits in which, at the same time as the comparison
population was generated, each of the 7 upper limits present in the data
was divided by some factor randomly drawn from a uniform distribution
between 1 and 50. The value 50 was chosen as being the ratio between
the highest jet prominence upper limit and the lowest actual
measurement, though we expect that the difference between the
upper limits and the true measurements will usually be less than
this. The effect is to increase the best-fit
values of $\beta_j$, but only slightly; this shows that the effect of
the upper limits on our results is not a serious one.

Whereas there is evidence from superluminal motions that relativistic
beaming should dominate the prominences of cores, we have little
direct evidence that it dominates the prominences of the kpc-scale
jets, as discussed in section \ref{intro}. It is thus worth considering
what effect random variation of intrinsic jet prominence (due to
`weather', as discussed above) might have on the best-fit beaming
speeds. In general this will reduce the best-fit $\beta_j$, because
less beaming is required to produce the observed spread in
prominences; clearly if the distribution of intrinsic prominence is
made to match the observed prominence distribution then no beaming is
necessary. A further constraint on prominence distributions is
provided, however, by the distribution of jet {\it sidednesses} in the
data; the sidedness of a source is defined, as in paper I, as the
ratio of the jet and counterjet prominences. If we assume that the
intrinsic prominences of the jet and counterjet are drawn from the
same distribution, then we can determine the distribution of
sidednesses in a simulated dataset. Because we only have a few
counterjet detections in the real dataset most of the sidedness data
we have come in the form of lower limits, while we have no constraints
at all on the sidedness of the 7 objects where no jet was detected. We
therefore incorporate the sidedness constraint by measuring the
fraction of the simulated sidednesses that match the measured ones,
either by being within $\pm 20$\% of them if they are actual
measurements or by being greater than them if they are upper limits. 7
of the simulated sidednesses are allowed to take any value. We then
multiply the K-S probabilities of a fit to the prominence data by this
fraction to give a (notional) total probability of matching the
data. It turns out that for $\beta_j \ga 0.5$ almost all the simulated
datasets can match the real dataset's sidedness in this sense; that
is, the weighting fraction is 1 for $\beta_j \ga 0.5$ and falls off
below it.

Unsurprisingly, then, when we draw the logs of the simulated intrinsic
prominences from a normal distribution, which is (as can be seen from
Fig. \ref{jetprom}) not badly matched to the data, allowing both the
mean ($p_j$) and standard deviation ($\sigma$) of the distribution
to vary, the best fits are achieved with $\beta_j \sim 0.5$, $\sigma
\sim 0.3$ (logs being taken to base 10). However, acceptable fits
(probabilities $>0.05$) are found for $\sigma \sim 0.4$, $\beta_j \la
0.1$. We cannot use the data on jet prominences and sidednesses alone
to rule out a model in which intrinsic jet and counterjet prominences
are drawn at random from a sufficiently broad distribution with {\it
no} or negligible relativistic beaming effects. If, however, we make
the (perhaps more physically reasonable) stipulation that intrinsic
jet and counterjet prominences are the same in particular sources,
then the simulations favour the relativistic beaming model more
strongly, and the best-fit $\beta_j$ is higher.

We note finally that it is likely that jets do not emit exactly
isotropically in their rest frames; this would affect the beaming
relation. Where good polarization measurements are available the
magnetic field in FRII jets appears to be directed longitudinally,
though it is likely, given that the polarizations of jets are
considerably less than the limiting value of $\sim 70$ per cent, that
we are actually seeing a projected, tangled field. In the limiting
case of a uniform longitudinal field the correct prominence relation
for jets is given by

\begin{equation}
p_j' = p_j [\gamma_j(1 - \beta_j \cos
\theta)]^{-(3+2\alpha)} (\sin \theta)^{1+\alpha}
\label{promdbpol}
\end{equation}
(e.g.\ Cawthorne 1991), producing a different expected distribution of
prominences for a given value of $\beta_j$. The increase in the exponent
of the Doppler factor and the additional term in $\sin \theta$ to some
extent cancel each other's effects. and so the cumulative probability
distributions are not actually very different over the whole possible
range of $\theta$ for the moderate values of
$\beta$ that seem to be appropriate here. When we performed our fits
using the prominence relation of equation \ref{promdbpol}, we obtained
very similar best-fit values of $\beta_j$ but slightly higher values of $p_j$.
If the magnetic field is not uniform and longitudinal, as expected,
then our analysis will be even less strongly affected.

\subsection{Relations}
\label{relation}

\begin{figure*}
\begin{center}
\leavevmode
\epsfxsize=13cm\epsfbox{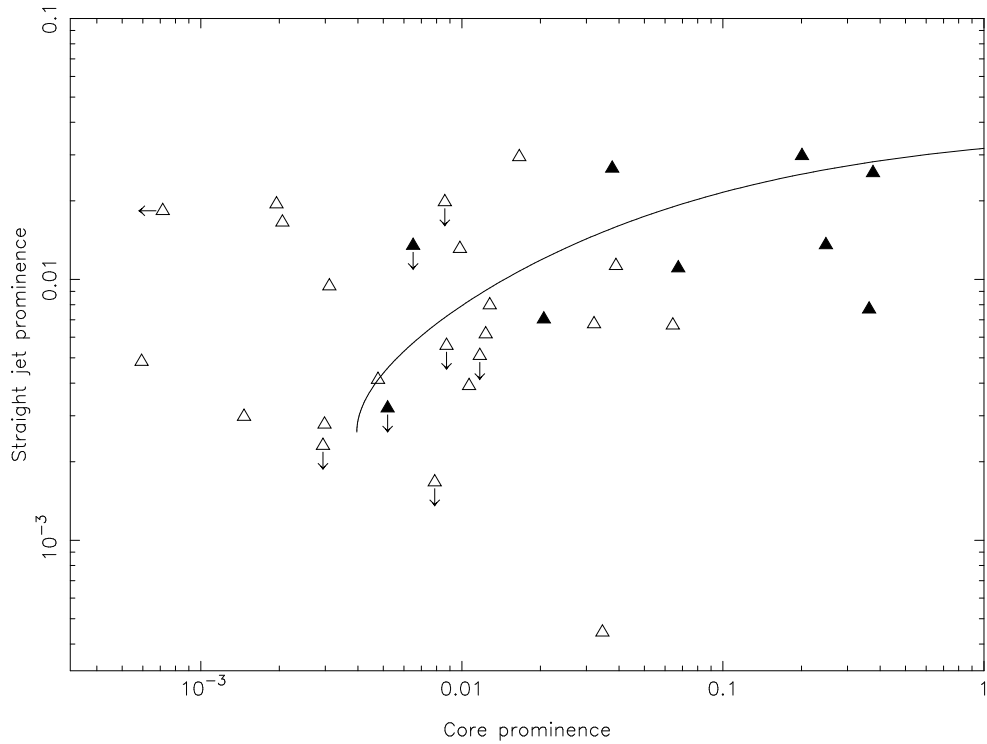}
\caption{The relationship between core and jet prominences for the
high-excitation objects in the sample. Filled symbols are BLRG, others
are NLRG; arrows indicate upper limits. The solid line shows a
theoretical curve assuming only relativistic beaming affects the
prominences, using $\beta_j = 0.62$, $p_{\rm j} = 0.005$, $\beta_c =
0.98$, $p_{\rm c} = 0.05$; see section \ref{relation}.}
\label{corejet}
\end{center}
\end{figure*}

In Fig.\ \ref{corejet} we plot the core and jet prominences for the 31
high-excitation objects in the sample. A correlation is apparent,
significant at the 95 per cent level on a modified Spearman Rank test
(as implemented in {\sc asurv}). As we argued in paper
I, if the prominences of cores are dominated by relativistic beaming,
such a correlation is not expected unless jets too are beamed; any
intrinsic relationship between core and jet would be washed out by the
beaming of the cores. But there
is clearly substantial deviation from the predictions of the simplest
relativistic beaming models, where intrinsic jet and core prominence
are the same for all sources; in such models the points in Fig.\
\ref{corejet} would be expected to lie on a single curved line (cf.\
Bridle \etal\ 1994) which we plot, for comparison, in Fig.\
\ref{corejet}. Scatter about this line can arise in a number of
ways. It is possible, for example, that there is normally jet bending
on 100-pc scales, so that the beaming angle $\theta$ is different for
core and jet; it is also possible that $\beta_c$ and $\beta_j$ vary
significantly from source to source. In the simplest models, though,
the scatter about the solid line in Fig.\ \ref{corejet} has
implications for the distribution of intrinsic core and jet
prominences, discussed above.

We have already argued that the cores are necessarily beamed, but that
the data seem to show a relationship between the intrinsic core
prominence and the linear size of the source --- a relationship which
is not obvious in the jet prominences (see Fig.\ 5 of paper I). We
have seen that jet prominences are well modelled with a combination of
intrinsic scatter and beaming. We may now ask whether these two
models, when combined, are consistent with the observed relationship
between core and jet; and we can try to make quantitative our
conclusion that models without beaming in the jets are ruled out by
this relationship.

We begin by asking whether the combined models are consistent with the
observed correlation found on the Spearman Rank test. Spearman's $r$
for the 31 real data points, as calculated taking limits into account
with {\sc asurv}, is 0.340. We simulate core and jet prominences based
on equations \ref{promdb2} and \ref{promdb} respectively, drawing the
logs of the intrinsic prominences of simulated jets from a normal
distribution as in section \ref{jets} and drawing the lengths of radio
sources from the distribution discussed in section \ref{cores}; for
cores, as before, we choose $\beta_c = 0.98$. Unsurprisingly, models
in which the spread in intrinsic prominence dominates the distribution
of jet prominences (so that jet and core are essentially unrelated)
cannot adequately reproduce the correlation; if $\beta_j = 0.1$,
$\sigma = 0.4$, the probability of obtaining a correlation at least as
strong as this is $\sim 5$ per cent, the level expected to occur by
chance, whereas if $\beta_j = 0.5$, $\sigma = 0.3$ it is a more
acceptable 75 per cent. We can rule out at the 90 per cent confidence
level all the best-fit models for jet prominence derived in section
\ref{jets} that have $\beta_j \la 0.2$, reinforcing our conclusion
that beaming is necessary in the jets.

As a final test, we can ask whether the distribution of core and jet
prominences in Fig.\ \ref{corejet} is consistent with the model we
have devised. Peacock (1983) provides a two-dimensional version of the
Kolmogorov-Smirnov test, subsequently refined by Fasano \& Franchesini
(1987); we use the implementation of their test given by Press \etal\
(1992), though we have verified that the original version of Peacock
(1983) gives similar results for our data. Fixing the parameters for
the core at $\beta_c = 0.98$, $p_{\rm c} = 0.05$, we allow the jet
$\beta_j$ and intrinsic prominence to vary; we find, as we expect,
that the results are similar to those plotted in Fig.\
\ref{jetks}. The data are (just) acceptably fitted by a model with no
intrinsic scatter, corresponding to the solid line in Fig.\
\ref{corejet}. However, they are better fitted (K-S probabilities
of 50--60 rather than 10 per cent) by models in which there is some
dispersion in the intrinsic prominences. If we allow the intrinsic
core prominences to depend on length, as above, and the intrinsic jet
prominences to be drawn from a normal distribution, then as before the
best results are obtained with $\beta_j \sim 0.5$, $\sigma \sim
0.3$. Models in which the jet and core are uncorrelated ($\theta_c$
and $\theta_j$ are both randomly chosen rather than being the same)
give acceptable but poorer fits to the data, while models in which the
jet and core are anticorrelated ($\theta_c = (\pi/2)-\theta_j$) are
never acceptable fits, as we would expect.

\section{Discussion and conclusions}
\label{discussion}

The results of the previous section allow us to conclude with little
ambiguity that relativistic beaming really is significant in
determining the prominences of the kpc-scale jets of these objects,
providing that it also dominates the prominences of the cores. Speeds
between 0.5 and $0.7c$ seem to provide the best fits to the
data. Speeds much higher than this do not reproduce the distribution
of jet prominences well, while lower speeds cannot well reproduce the
observed relationships between core and jet prominence and the jet
sidedness distribution, and are in any event at odds with observations
of the Laing-Garrington effect. These conclusions are similar to those
in earlier work (Bridle \etal\ 1994; Hough 1994; Wardle \& Aaron 1997)
but they are based on a sample that should be free of (unknown)
orientation bias. Subject to the caveats already discussed, the data
and the relativistic beaming speeds derived from them are broadly
consistent with the simple unified model of paper I, in which BLRG are
aligned at angles less than $\sim 45\dgr$ to the line of sight.

Are these beaming speeds a reflection of any real physical
velocity in the beams? As pointed out by Bridle \etal\ (1994), it is
entirely possible that the material in the beam does not move with a
single speed. If this is true, the speed derived from a {\it
single} object is, as they say, a weighted average over the velocity
field in the beam; the speed derived by application of equation
\ref{promdb} to a whole sample of beamed objects is still more
complicated in its relationship to any true physical
speeds. Related to this is the whole question of where the jet
emission comes from, and why, in fact, we see emission from the beam
at all. However, if (as has been suggested; e.g.\ Bridle 1996) the
emission in these objects comes from a shear layer at the edge of the
beam, while the emission from a higher-velocity central spine is
suppressed to negligible levels by beaming, then the characteristic
speeds we derive may well be a reasonable estimator of the
physical speeds in the shear layer. Such a picture may also help
us to understand why the speeds derived from cores and jets are so
different; perhaps we are simply seeing different velocity regimes
because different regions of the beam are emitting. Alternatively, it
may be that the difference is evidence of a genuine deceleration of
the beam between the two scales. Such deceleration is invoked in
models of FRI radio sources, where similar speeds around $0.5c$
are derived, from sidedness arguments, at the bases of the jets;
[e.g.\ Hardcastle \etal\ (1996), Laing (1996), Hardcastle \etal\ (1997)].
As discussed by Bowman, Leahy \& Komissarov (1996) it is possible to
decelerate FRII beams by entrainment without excessive dissipation.

Another, related question is raised by the derived intrinsic
prominences. The best-fit intrinsic prominences of cores in our sample
are $\ga 0.02$, with the value for our representative $\beta_c = 0.98$
($\gamma_c \approx 5$) being about 0.05. These numbers are similar to
those that have been derived in similar ways from other samples
[e.g. Morganti \etal\ (1995) and references therein]. But the best-fit
intrinsic prominences of jets are lower; $p_j \approx 0.005$ is a
typical value derived from the fits for $\beta_j \approx 0.6$. In
other words, even if the sources were not beamed, the sub-kpc-scale
radio emission from the cores would be brighter than that from the
much larger 100-kpc-scale jets. The beam is evidently much less
efficient during the first few hundred pc of its journey to the hot
spots than it is at any future time. This may be a sign either of
strong interaction with the external medium or of shocks in the jet
on these small scales, which would be consistent with a model in which
the beam is decelerating strongly at this stage.

It is worth noting that the efficiency of the jet in these FRIIs is
still very high; because the total radio luminosity of an extended
source is much lower than the total power supplied by the beam (cf.\
Rawlings \& Saunders 1991) and the unbeamed luminosity of the core and
jet is less, by approximately $p_j + p_c$, than the total luminosity
of the source, it follows that the efficiency of the jet is $\gg [1 -
(p_j + p_c)]$, or probably $>99$ per cent. Almost all the energy
supplied by the beam is used to excite the lobe electron population
and to do work on the external medium, and very little of it is
radiated away in jets.

\section*{Acknowledgements}

MJH acknowledges a studentship from PPARC and subsequent support from
PPARC grant GR/K98582. We thank participants at the 39th Herstmonceux
symposium, particularly Robert Laing and Tim Cawthorne, for helpful
comments on a version of this work presented there, and an anonymous
referee for suggestions that helped us to improve the clarity of the
paper. The National Radio Astronomy Observatory is operated by
Associated Universities Inc., under co-operative agreement with the
National Science Foundation.

\bsp

\begin{table*}
\caption{Fluxes and prominences for the sample sources}
\label{numbers}
\begin{tabular}{ld{2}ld{3}d{3}d{3}d{3}d{5}d{3}d{3}d{3}d{3}d{5}}
&&&&&&&&\multicolumn{2}{c}{North jet}&\multicolumn{2}{c}{South jet}\\
\multicolumn{1}{l}{Source}&\multicolumn{1}{c}{Freq.}&\multicolumn{1}{c}{Line}&\multicolumn{1}{c}{Flux}&\multicolumn{1}{c}{Error}&\multicolumn{1}{c}{Core flux}&\multicolumn{1}{c}{Error}&\multicolumn{1}{c}{Core}&\multicolumn{1}{c}{Flux}&\multicolumn{1}{c}{Error}&\multicolumn{1}{c}{Flux}&\multicolumn{1}{c}{Error}&\multicolumn{1}{c}{Brighter jet}\\
&\multicolumn{1}{c}{(GHz)}&\multicolumn{1}{c}{class}&\multicolumn{1}{c}{(Jy)}&&\multicolumn{1}{c}{(mJy)}&&\multicolumn{1}{c}{prom.}&\multicolumn{1}{c}{(mJy)}&&\multicolumn{1}{c}{(mJy)}&&\multicolumn{1}{c}{prominence}\\
\hline
\Ss{3C15}&8.35&E&1.00&&28.00&0.05&0.04&96&3&<1.4&&0.1\\
\Ss{3C20}&8.44&N&2.29&&3.32&0.06&0.001&7&5&<9&&0.003\\
\Ss{3C33.1}&1.53&B&3.02&&20.40&0.07&0.02&<37&&27&4&0.007\\
\Ss{3C79}&8.44&N&0.694&&6.04&0.01&0.009&<3.8&&<1.3&&<0.006\\
\Ss{3C98}&8.35&N&3.08&0.07&6.1&0.1&0.002&50&20&<13&&0.02\\
\Ss{3C105}&8.35&N&1.68&&18.9&0.5&0.01&<8.3&&<48&&<0.005\\
\Ss{4C14.11}&8.44&E&0.500&&29.69&0.03&0.06&<0.45&&1.14&0.04&0.002\\
\Ss{3C111}&8.35&B&4.8&0.2&1276&1&0.4&90&20&<58&&0.03\\
\Ss{3C123}&8.44&E&9.44&&108.9&0.3&0.01&<2.1&&<30&&<0.003\\
\Ss{3C132}&8.44&E&0.674&&4.1&0.2&0.006&<13&&2&3&0.004\\
\Ss{3C135}&8.35&N&0.520&&1.0&0.2&0.002&<1.6&&10&4&0.02\\
\Ss{3C136.1}&8.35&?&1.00&0.05&1.53&0.03&0.002&<2.6&&<2.7&&<0.003\\
\Ss{3C153}&8.44&N&0.712&&<0.5&&<0.0007&8&1&13&5&0.02\\
\Ss{3C171}&8.06&N&0.690&&2.0&0.1&0.003&6.0&0.7&6.3&0.8&0.009\\
\Ss{3C173.1}&8.44&E&0.461&&9.64&0.02&0.02&2.1&0.9&<0.29&&0.005\\
\Ss{3C184.1}&8.35&N&0.785&&6.0&0.5&0.008&<14&&<1.3&&<0.002\\
\Ss{3C192}&8.35&N&1.38&0.08&4.0&0.2&0.003&<3.2&&<7.7&&<0.002\\
\Ss{3C197.1}&8.35&E&0.320&&6.0&0.1&0.02&<1.9&&<0.85&&<0.006\\
\Ss{3C219}&4.87&B&2.27&0.06&51.6&0.1&0.04&56.5&0.3&2.1&0.1&0.03\\
\Ss{3C223}&8.35&N&0.89&0.05&8.5&0.2&0.01&11&4&<5.8&&0.01\\
\Ss{3C223.1}&8.35&N&0.53&0.01&6.4&0.4&0.01&3&1&<21&&0.006\\
\Ss{3C227}&8.35&B&2.05&0.04&13.2&0.6&0.007&<15&&<27&&<0.01\\
\Ss{3C234}&8.44&N&0.919&&34.46&0.04&0.04&10&8&<19&&0.01\\
\Ss{3C284}&8.06&N&0.340&&2.79&0.02&0.009&<2.8&&<6.6&&<0.02\\
\Ss{3C285}&4.86&N&0.740&&6.8&0.4&0.02&19&2&<14&&0.03\\
\Ss{3C300}&8.06&N&0.645&&6.2&0.1&0.01&2.4&0.2&<0.23&&0.004\\
\Ss{3C303}&1.48&B&2.45&&106.6&0.3&0.2&63&5&<13&&0.03\\
\Ss{3C319}&8.44&E&0.362&&<0.3&&<0.0008&<0.12&&<1.8&&<0.005\\
\Ss{3C327}&8.35&N&2.01&0.05&25&1&0.01&16&6&<140&&0.008\\
\Ss{3C349}&8.44&N&0.723&&24.21&0.02&0.03&<0.053&&0.31&0.04&0.0004\\
\Ss{3C353}&8.44&E&14.1&&151.0&0.2&0.01&70&10&<34&&0.005\\
\Ss{3C381}&8.44&B&0.906&&4.7&0.1&0.005&<2.9&&<1.3&&<0.003\\
\Ss{3C382}&8.35&B&1.30&&251.2&0.1&0.2&14&1&<1.1&&0.01\\
\Ss{3C388}&4.87&E&1.80&&57.9&0.1&0.05&30&5&34&6&0.02\\
\Ss{3C390.3}&8.35&B&2.8&0.1&733&5&0.4&20&10&<650&&0.008\\
\Ss{3C401}&8.44&E&0.844&&28.54&0.03&0.04&<5.7&&33.8&0.4&0.05\\
\Ss{3C403}&8.35&N&1.50&&7.1&0.2&0.005&6&1&<3.8&&0.004\\
\Ss{3C405}&4.53&N&415&&776&3&0.003&1200&600&500&800&0.003\\
\Ss{3C424}&8.35&E&0.357&&7.0&0.3&0.02&<8.4&&16.7&0.9&0.05\\
\Ss{3C433}&8.35&N&2.08&&1.2&0.3&0.0006&9.8&0.1&<8.3&&0.005\\
\Ss{3C436}&8.44&N&0.592&&17.90&0.02&0.03&<0.27&&3.8&0.8&0.007\\
\Ss{3C438}&8.44&E&0.780&&16.2&0.1&0.02&40&4&<9.8&&0.06\\
\Ss{3C445}&8.40&B&1.34&0.08&83.9&0.4&0.07&<1.9&&14&3&0.01\\
\Ss{3C452}&8.35&N&2.14&&125.8&0.3&0.06&9&2&13&2&0.007\\

\end{tabular}
\parbox{\linewidth}{Column 2 gives the observing frequency of the
radio data used in the analysis. Prominences are corrected to a
nominal frequency of 8.4 GHz as described in paper I. Column 3 lists
the emission line type of the source. `E' indicates a LERG, `N' a NLRG
and `B' a BLRG; '?' indicates an unclassified source. Paper I provides
the references for these classifications. Column 4 gives the total
flux density at the observing frequency. Errors on this quantity (column 5)
are quoted only where the flux density was taken from the literature rather
than VLA maps, or where there was some uncertainty in the definition
of the source region on the VLA maps, as discussed in paper I; the
standard VLA calibration errors apply to all other measured quantities.}
\end{table*}

\end{document}